\documentclass[prl,aps,twocolumn,showpacs,superscriptaddress,dvipsnames]{revtex4-1}

\usepackage{amsmath,amssymb,bbm,mathtools}
\usepackage{graphics,epsfig,psfrag,xcolor}
\usepackage[colorlinks=true,citecolor=MidnightBlue,linkcolor=MidnightBlue,urlcolor=MidnightBlue]{hyperref}

\usepackage{tgtermes}
\usepackage[T1]{fontenc}
\usepackage[scaled]{helvet}

\newcommand{\aperture}{\mathcal{A}}
\newcommand{\ket}[2][]{\lvert #2\rangle_{#1}}
\newcommand{\bra}[2][]{{}_{#1} \langle #2\rvert}
\newcommand{\braket}[1]{\langle #1 \rangle}
\newcommand{\LeftTraceBracket}{[}%
\newcommand{\RightTraceBracket}{]}%
\newcommand{\tr}[2][]{\mathrm{Tr}_{#1} \LeftTraceBracket #2 \RightTraceBracket}
\newcommand{\Tr}[2][]{\mathrm{Tr}_{#1} \mathopen{}\left\LeftTraceBracket #2 \right\RightTraceBracket\mathclose{}}
\newcommand{\pr}[2][]{\mathrm{Pr}_{#1} ( #2 )}
\newcommand{\abs}[2][]{\lvert#2\rvert_{\text{#1}}}
\newcommand{\Abs}[2][]{\left\lvert#2\right\rvert_{\text{#1}}}
\newcommand{\identity}{\mathbbm{I}} % 

\begin{document}

\title{{Fundamental Limits of Classical and Quantum Imaging} } 

\author{Carlos A. P\'erez-Delgado}%\email{caperez@gmail.com}
\affiliation{Centre for Quantum Technologies, National University of Singapore,  3 Science Drive 2, Singapore 117543, Singapore}
\affiliation{Department of Physics \& Astronomy, the University of Sheffield, Hounsfield road, Sheffield, S3 7RH, UK}  

\author{Mark E. Pearce}%\email{mrkprc1@gmail.com}
\affiliation{Department of Physics \& Astronomy, the University of Sheffield, Hounsfield road, Sheffield, S3 7RH, UK}  

\author{Pieter Kok}\email{p.kok@sheffield.ac.uk}
\affiliation{Department of Physics \& Astronomy, the University of Sheffield, Hounsfield road, Sheffield, S3 7RH, UK}

\begin{abstract}
\noindent 
Quantum imaging promises increased imaging performance over classical protocols. However, there are a number of aspects of quantum imaging that are not well understood. In particular, it has so far been unknown how to compare classical and quantum imaging procedures. Here, we consider classical and quantum imaging in a single theoretical framework and present {general fundamental limits on the resolution and the deposition rate} for classical and quantum imaging. The resolution can be estimated from the image itself. We present a utility function that allows us to compare imaging protocols in a wide range of applications. 
\end{abstract}


\date{\today}
\maketitle

\noindent
Imaging is an important technological tool in many disciplines including astronomy, biomedical research, nanotechnology and basic  research. The wave nature of light limits the resolution and contrast that can be achieved in classical imaging techniques. On the other hand, quantum entanglement may offer some improvement over these classical limits, leading to the subject of quantum imaging. The best known quantum imaging protocols are two-photon microscopy \cite{fei97,teich97,abouraddy01} and spectroscopy \cite{agarwal01}, quantum holography \cite{sokolov01}, quantum lithography \cite{boto00}, and quantum illumination \cite{lloyd08}. The subject of quantum imaging is closely related to quantum metrology \cite{kok04}, which is quite well understood \cite{giovannetti06,Zwierz10}. However, there are some important differences. If we restrict our discussion to quantum optics, quantum metrology is first and foremost concerned with photon statistics, which is governed almost entirely by transformations of the mode 
operators of the quantum field. Quantum imaging, on the other hand, is intimately related to both the operator transformations and the specific form of the (classical) mode functions that encode the spatial properties of the field. This has made it difficult to identify the essential quantum mechanical behavior in imaging protocols.

There are two reasons why it is important to understand the precise distinction between classical and quantum imaging. First, it can help identify new methods for improved imaging, potentially leading to new technologies. Second, it may reveal a fundamental aspect of physics that has hitherto remained elusive, namely what (if any) makes quantum optics more powerful for imaging than classical optics. The obvious answer to this question---quantum entanglement---has already been proved false to some extent \cite{tsang08}. While entanglement is likely necessary for an improved imaging procedure, it is certainly not sufficient. This is reminiscent of quantum computing, where it was shown that entanglement is necessary but not sufficient for obtaining the promised exponential speed-up over classical computing \cite{gottesman97}.

Here we develop a theoretical framework that allows us to study classical and quantum imaging in a unified fashion, { which carries over tools and techniques from the mathematical theory of metrology to prove fundamental limits. This is fundamentally and conceptually different from studying parameter estimation systems that happen to use light as the probing mechanism} (see e.g., \cite{delaubert08}). Our methodology is based on the statistical distance between two probability distributions that characterise two images. We find that the distinguishability of the probability distributions provides a natural definition of the imaging resolution in terms of a Cram\'er-Rao bound. We apply this bound to several imaging procedures and show that it gives the desired result. We define a utility function that provides a general metric for imaging procedures in practical applications.

\begin{figure}[t]
 \centering
 \includegraphics[width=6.5cm]{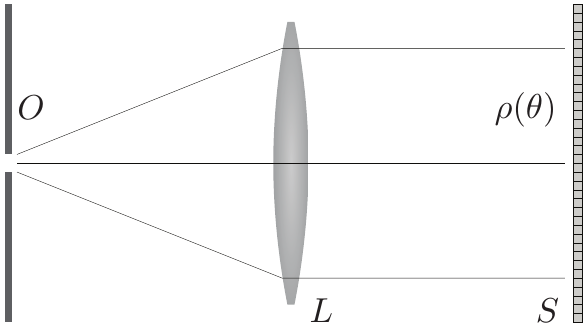}
 \caption{General imaging procedure. The state of light $\rho(\theta)$ contains information $\theta$ about the object $O$ and the imaging system $L$. The substrate $S$ acts as a measurement of $\rho(\theta)$ and is generally described by a {\sc povm}. This model includes multi-photon quantum imaging procedures.}
 \label{fig:imaging}
\end{figure}

The most general imaging procedure is shown in Fig.~\ref{fig:imaging}. Light emitted or reflected by an object $O$ is optically processed by an imaging system $L$ and recorded by a substrate $S$. Typically, the substrate has a granular spatial structure so we can break it up into $N$ systems indicated by their position $x$. If the width of such a ``pixel'' is $\ell$, then the total length of the substrate is $L = N\ell$. Moreover, we assume that the intensity is granular (which is consistent with the use of photons at the fundamental level), and the values of the image intensity at a pixel are given by $i_k$. The average recorded intensity at pixel $x$ is then given by $I(x) = \sum_k i_k\, p_k(x)$, where $p_k(x)$ is the probability of measuring $i_k$ at pixel $x$. In quantum mechanics, this is given by the Born rule $p_k(x) = \tr { \rho \hat{E}_k(x)}$, with $\rho$ the quantum state of light incident on the substrate and $\hat{E}_k(x)$ the measurement operator for outcome $i_k$ at $x$. The set of $\hat{E}_k(x)
$ form a {positive operator-valued measure ({\sc povm})} for each pixel $x$. The image $I(x)$ may be related in a simple manner to the intensity of the light that hits the substrate, or it may have a more complicated relationship involving multi-photon absorption, etc. For monochromatic images, $I(x)$ will be a single-valued function indicating the brightness of the image at $x$. Note that the pixels do not need to be placed side-by-side (see, e.g. Ref.~\cite{oppel12}). For simplicity we will limit our discussion to one-dimensional substrates in the remainder; however all our results generalise easily.

We can construct the {image observable} $\hat{I}(x)$ that is measured by the substrate. We define $I(x) = \braket{\hat{I}(x)}$ such that 
$I(x) = \sum_k i_k\, \tr{\rho \hat{E}_k(x) } \equiv \tr { \rho\, \hat{I}(x)}$, and we therefore find that $\hat{I}(x) = \sum_k i_k\, \hat{E}_k (x)$. In the simplest case where the substrate records the number of photons $k$, the {\sc povm} elements are $\hat{E}_k(x) = \ket[x]{k}\bra{k}$ and $i_k = k$. More complicated {\sc povm}s can incorporate for example photon losses, a saturation point, bleeding from adjacent points, {or multi-photon interference}. In typical imaging protocols the measurement at each pixel is diagonal in the Fock basis, and the properties of the substrate are characterised by the {\sc povm}
\begin{align}\label{eq:povm}
 \hat{E}_k(x) = \sum_{\vec{n}} q_k(x|\vec{n}) \ket{\vec{n}}\bra{\vec{n}}\, ,
\end{align}
where $q_k(x|\vec{n})$ is the conditional probability of finding measurement outcome $k$ at pixel $x$ given the photon number distribution $\vec{n} = (n_1,\ldots,n_N)$ over $N$ pixels, and $\sum_k q_k(x|\vec{n}) = 1$. We identify the recorded {image} $I(x)$ with an unnormalized probability distribution. After normalisation we define 
\begin{align}\label{eq:prob}
 \pr{x} \equiv \frac{I(x)}{I_0}\, ,
\end{align}
where {$I_0 = \sum_x I(x)$ is the image normalisation}. 

In imaging, we are often interested in distinguishing two possibilities, which we can para\-me\-trize by a continuous number $\theta$, for example the position of a light source. Hence, the probability distribution $\pr{x|\theta}$, and therefore the image, will depend on $\theta$. We now have two probability distributions, $\pr{x|\theta=0}$ and $\pr{x|\theta}$. Distinguishing two images is then possible if we can distinguish the two probability distributions. One method to achieve this is to define the infinitesimal statistical distance \cite{braunstein94}
\begin{align}
 ds^2 =  \sum_x \frac{[\pr{x|d\theta}-\pr{x|0}]^2}{\pr{x|0}}\, ,
\end{align}
which can be formally integrated to a function $s(\theta)$. {We note, however, that in principle \emph{any} method for distinguishing probability distributions can be used.} Using a Taylor expansion of $s(\theta)$ up to first order in $\delta\theta$, we can write 
\begin{align}
 \delta s \equiv s(\delta\theta) - s(0) = \left. \frac{ds}{d\theta} \right|_{\theta=0} \delta\theta\, .
\end{align}
The two probability distributions are distinguishable when the statistical distance between them exceeds a constant value $(\delta s)^2 \geq 1$, which we can rewrite as 
\begin{align}\label{eq:distcritF}
 (\delta s)^2 = F(0) (\delta\theta)^2\geq 1\qquad\text{with}\quad F(\theta) = \left( \frac{ds}{d\theta} \right)^2\, .
\end{align}
The quantity $F(\theta)$ is known as the Fisher information \cite{helstrom76}
\begin{align}
 F(\theta) = \sum_x \frac{1}{\pr{x|\theta}} \left( \frac{d \pr{x|\theta}}{d\theta} \right)^2 ,
\end{align}
and is a measure of the information we can extract about $\theta$ given the probability distribution $\pr{x|\theta}$. When we use the form of $\text{Pr}(x|\theta)$  in Eq.~\eqref{eq:prob} we find  
\begin{align}\label{eq:Fisher2}
F(\theta) = \sum_x \frac{[\partial_{\theta} \braket{\hat{I}(x)}]^2}{I_0 \braket{\hat{I}(x)}} - \frac{(\partial_{\theta}I_0)^2}{I_0^2},
\end{align}
where $\partial_{\theta} \equiv \partial/\partial\theta$. If $I_0$ is independent of $\theta$ the last term vanishes. The Fisher information is therefore calculated directly from the recorded image, without the need for modelling the {\sc povm} $\{ \hat{E}_k(x)\}$ of a complicated substrate. 

We can translate the distinguishibility criterion in Eq.~(\ref{eq:distcritF}) using $F_0 \equiv F(0)$ into a \emph{resolution limit} for the parameter $\theta$ and find that 
%$\delta\theta \geq \smash{F_0^{-\frac12}}$.
\begin{align}
 \delta\theta \geq \frac{1}{\sqrt{F_0}}\, .
\end{align}
This is a single-shot Cram\'er-Rao bound for the imaging resolution of $\theta$, which is achievable in the limit of a well-exposed image. This result is readily extended to multiple parameters $\bm{\theta} = (\theta_1,\ldots,\theta_D)$ using well-known techniques \cite{helstrom76}. 

There is a key difference between the theory of imaging presented here and the theory of quantum metrology \cite{giovannetti11}. In the latter, the Cram\'er-Rao inequality gives a bound on the mean square error in the parameter $\theta$ as a function of the number of independently repeated measurements \cite{delaubert08}. By contrast, in quantum imaging the Cram\'er-Rao bound is based on the full probability distribution $\pr{x|\theta}$, and hence assumes that the image is sufficiently well-exposed such that Eq.~(\ref{eq:prob}) holds. Faint images can be modelled phenomenologically by adding a (uniform) probability distribution to Eq.~(\ref{eq:prob}).

Next, we demonstrate that this technique gives the correct results for some known imaging procedures. First, we consider the double slit experiment, in which the image is the familiar interference pattern in the far field \cite{lipson11}. We parameterise the slit separation by $\theta$ and have the numerical aperture $\aperture$ capture a fraction of the interference pattern.  To calculate the Fisher information we use the full form of Eq.~(\ref{eq:Fisher2}) since $I_0$ will be heavily dependent on $\theta$.  We then identify that the resolution criterion is that the uncertainty $\delta\theta$ is smaller than the estimate $\theta$.  From this we derive that two slits are resolved when $\theta^2F(\theta) \geq 1$.  We then calculate the resolution limit to be $\theta_{\rm min}\geq 0.369 \lambda/\aperture$.  When we compare this to the Abbe limit $\theta_{\rm min} \geq 0.5 \lambda/\aperture$, we find a slightly better resolution.  This is because the Abbe limit is a conservative estimate that does not take into 
account the extra information about $\theta$ in the slopes of the intensity pattern. The Fisher information is very sensitive to these slopes. {When we calculate the resolution for the third-order photon interference slit imaging experiment by Oppel et al., \cite{oppel12}, we find the factor 2 improvement, as expected.}

\begin{figure}[t]
\includegraphics[width=8.5cm]{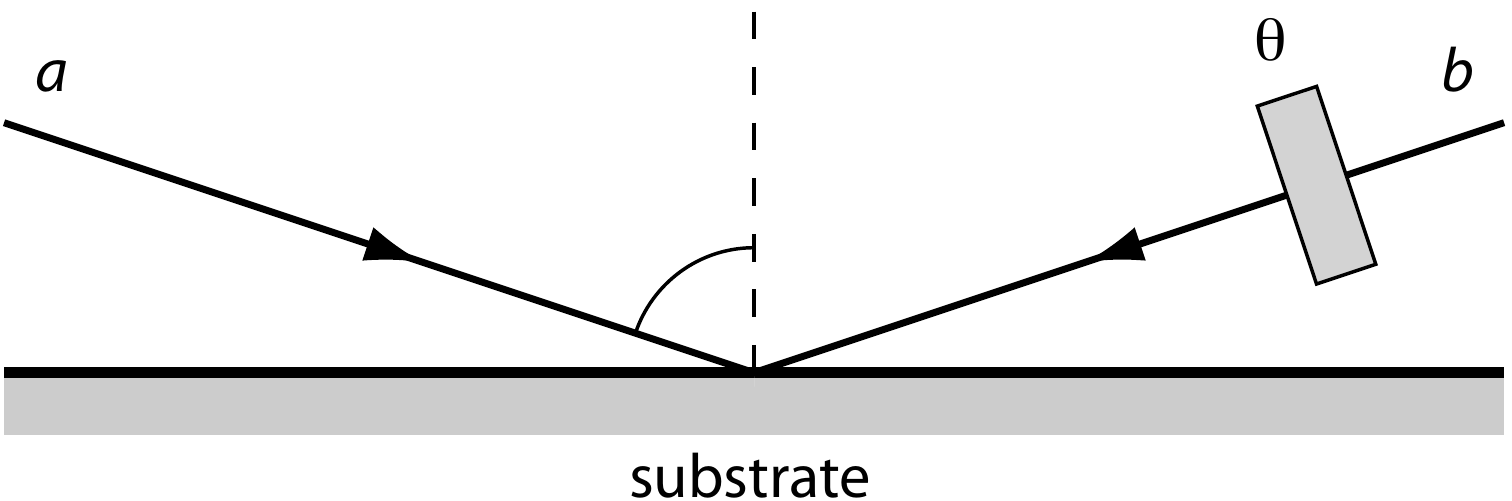}
\caption{Optical lithography. Two nearly counter-propagating beams $a$ and $b$ produce an interference pattern on the substrate. The phase shift $\theta$ causes a shift in the image.}
\label{fig:litho}
\end{figure}

\begin{figure}[t]
\includegraphics[width=8.5cm]{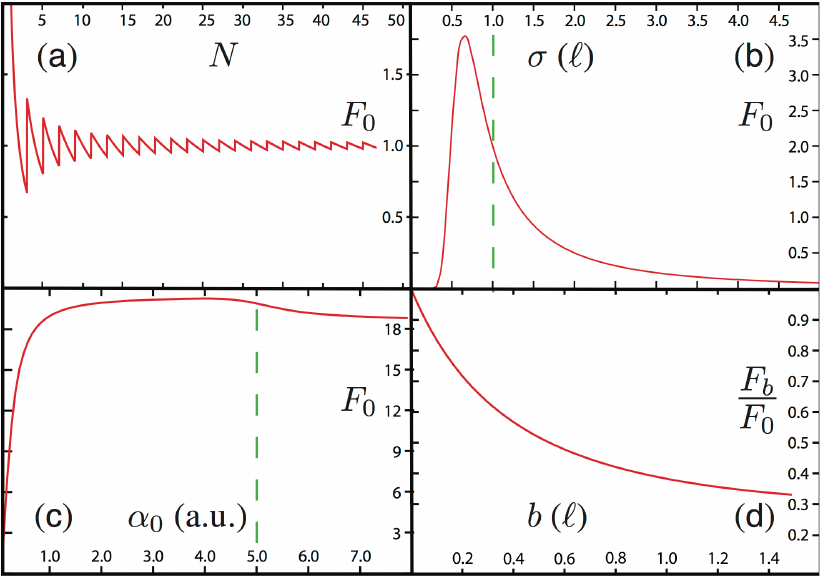}
\caption{(color online) Qualitative behaviour of the Fisher information: ({\sf a}) $F_0$ in Eq.~(\ref{eq:litho}) as a function of the number of pixels $N$ tends to 1 for large $N$; ({\sf b}) $F_0$ for the displacement of a Gaussian dot with width $\sigma$ (in units of pixel size $\ell$). The vertical dotted line indicates where $\sigma=\ell$, the size of one pixel; ({\sf c}) $F_0$ for a Gaussian dot as a function of the peak amplitude $\alpha_0$ (in arbitrary units). The vertical doted line indicates the saturation point at  $\alpha_0 = 5$, {and $\sigma = 70\ell$}. The saturation point has only a small effect, since $F_0$ is sensitive mostly to intensity slopes; ({\sf d}) The ratio of the Fisher information with and without bleeding $F_b/F_0$ as a function of the average bleeding distance $b$ {(with $\sigma = 40\ell$ and $\alpha_0 = 70$)}.}
\label{fig:Fisher}
\end{figure}

Second, we consider classical and quantum photo-litho\-graphy \cite{boto00} shown in Fig.~\ref{fig:litho}. In the classical regime the input state can be modelled by the single photon superposition state $\ket{\psi} = (\ket[ab]{1,0} + e^{i\theta} \ket[ab]{0,1})/\sqrt{2}$, where $\theta = \kappa\ell\Delta x$ shifts the pattern along the substrate ($\kappa = 2\pi/\lambda$ is the wave vector). {We give the {\sc povm} of a general $M$-photon absorbtion substrate including a detection efficiency $\eta$. The conditional probabilities in Eq.~(\ref{eq:povm}) are $q_0(x|n_x) = 1$ for $n_x < M$ and $q_0(x|n_x) = (1-\eta)^{n_x}$ for $n_x \geq M$; $q_1(x|n_x) = 1-q_0(x|n_x)$. The Fisher information for $M=1$} is then calculated as
\begin{align}\label{eq:litho}
F(\theta) = \frac{2}{N}\sum_x \text{sin}^2\left(\kappa lx+\frac{\theta}{2}\right)\leq2\, .
\end{align}
If the substrate is much larger than the wavelength of the light, then $F_0 \simeq 1$ (see Fig.~\ref{fig:Fisher}a) and the Cramer-Rao bound is $\delta\theta \geq 1$, or $\ell\delta x \geq \lambda/2\pi$. This imaging resolution is slightly better than the Rayleigh criterion $\ell\delta x \geq \lambda/4$ due to the slope sensitivity of the Fisher information.

For $M$-photon quantum lithography the Fisher information is given by
\begin{align}
 F(\theta) = \frac{2M^2}{N}\sum_x \text{sin}^2\left(M\kappa lx+\frac{M\theta}{2}\right).
\end{align}
When $\kappa lN \gg 1$ the Fisher information is $F_0 \simeq M^2$, yielding $\delta\theta \geq M^{-1}$. The imaging resolution $\ell\delta x \geq \lambda/2\pi M$ is slightly better than $\ell\delta x \geq \lambda/4M$ found by Boto \emph{et al}.~\cite{boto00}. In both the classical and quantum case the resolution is independent of the efficiency $\eta$.  

Finally, we consider imaging the position of a dot of coherent light with a Gaussian intensity profile on a one-dimensional substrate. We include a saturation limit of the substrate and pixel bleeding due to interactions between adjacent pixels. For simplicity the state of light on the substrate is given by $\ket{\bm{\alpha}}=\bigotimes_{x=1}^N \ket[x]{\alpha_x}$ with $\ket[x]{\alpha_x}$ a coherent state of amplitude $\alpha_x = \alpha_0\,\text{exp}[-(x-x_0)^2 /2\sigma^2]$ at pixel $x$. There is no coherence between adjacent pixels. Assuming $\sigma\ll L$ and $x_0$ towards the centre of the substrate (so $\partial_{\theta} I_0 \approx 0$), the Fisher information for perfect photodetection at each pixel becomes
\begin{align}
F(\theta) = 4 \sum_x \frac{\Abs{\alpha_x}^2}{I_0} \frac{(x-\theta)}{\sigma^2}\, .
\end{align}
The Fisher information is shown as a function of the standard deviation $\sigma$ in Fig.~\ref{fig:Fisher}b. When $\sigma$ falls below one pixel width the Fisher information drops rapidly to zero.  This is expected since below one pixel width we are no longer able to detect a correspondingly small shift in $\theta$.  Above one pixel width the graph follows the form of $F = 2/\sigma^2$, which gives the resolution $\delta\theta \geq \sigma/\sqrt{2}$.

Taking into account the saturation limit $S$ leads to conditional probabilities in the {\sc povm} given by
\begin{align}\label{eq:satprob}
q_k(x|\vec{n}) = \left\{
\begin{array}{l l}
\delta_{k,n_x} & \quad \text{if $k < S$}, \\
\sum_{j=0}^{\infty} \delta_{n_x,S+j} & \quad \text{if $k = S$}. \\
\end{array}
\right.
\end{align} 
For the displacement of a Gaussian dot this gives the Fisher information shown in Fig.~\ref{fig:Fisher}c. Due to a uniform noise floor $F_0$ rises sharply for small $\alpha_0$ and falls away at the saturation point.  The noise is numerically necessary to avoid near-singularities. {The saturation truncates relevant statistical information, but the main contribution to $F_0$ comes from the steep ascent regions of the Gaussian, rather than the crown. Therefore, the overall impact on the Fisher information is relatively small for this particular imaging process. In other situations, the saturation limit can have severe implications for the resolution.}

When there exists an interaction between adjacent pixels, a signal at pixel position $x$ is recorded by its neighbours with some probability. This is called \emph{bleeding}, and the conditional probabilities of Eq.~\eqref{eq:povm} are given by $q_k(x|\vec{n}) = \sum_{x^{\prime}}\delta_{k,n_{x^{\prime}}}\pr{d}$, where $\pr{d}$ is the probability of a signal bleeding a distance $d = \abs{x^{\prime}-x}$.  For a Poisson distribution the average distance is characterised by the bleeding parameter $b$. The qualitative effect on the Fisher information is shown in Fig.~\ref{fig:Fisher}d. 

Given the Cram\'er-Rao bound for the imaging resolution, we would like to find a physical limit to the Fisher information that in turn would give a fundamental limit to all imaging procedures. The Fisher information is bounded by $F( \theta ) \leq 4(\Delta K )^2/\hbar^2$ \cite{braunstein96b},
%\begin{align} 
% F( \theta ) \leq \frac{4}{\hbar^2} ( \Delta K )^2\, ,
%\end{align}
where $\Delta K$ is the variance of $K$, the generator of translations in $\theta$ such that $\rho(\theta) = U \rho(0) U^{\dagger}$ with $U=\exp(-i\theta K/\hbar)$. However, this relation does not include any information about the substrate, since it is a general bound on any possible metrological system including ideal imaging procedures.  Another important question therefore concerns the best possible resolution over all imaging systems and substrates given a particular constraint on the substrate. For example, what is the best imaging resolution given a certain saturation limit $S$, or a given amount of pixel bleeding $b$? This is currently an open question.

As pointed out by Tsang \cite{tsang07} and Steuernagel \cite{steuernagel04}, the practicality of some quantum imaging procedures such as quantum lithography \cite{boto00} is severely affected by the efficiency of the measurement procedure. In the case of $M$-photon lithography the rate of the $M$-photon detection events at pixel $x$ yields our image $I(x)$. This is calculated as
\begin{align}
 I(x) = 2\left( \frac{\eta}{2N} \right)^M \cos^2\left( M\kappa \ell x + M\frac{\theta}{2} \right) .
\end{align}
The constant of proportionality disappears in the normalisation procedure of Eq.~(\ref{eq:prob}), and hence does not contribute to the Fisher information. However, it is clear that the \emph{deposition rate} of $M$-photon detection events falls off exponentially. Any practical figure of merit must therefore take into account both the resolution and the deposition rate of the imaging procedure. We define a \emph{utility function} $U$ for an imaging procedure {as $U = F_0 D^c$, which} should be maximised. Here, $0\leq D\leq 1$ is the deposition rate and $c > 0$ is a cost function that can take into account the resource cost of each attempted detection event. This may reflect both physical and economical aspects of the imaging procedure. For example, in the manufacturing of microchips speed is an important factor and the cost function should be suitably weighted to reflect this. The deposition rate can be defined based on the {\sc povm} element $\hat{E}_0$ that indicates a failure to detect anything. If $\hat{E}
_0(x)$ is the {\sc povm} element for not detecting anything at pixel $x$, then $\hat{E}_0 = \bigotimes_{x=1}^N \hat{E}_0(x)$ and the deposition rate is given by 
\begin{align}
 D = \sum_{x=1}^N \Tr{\rho(\theta)(\identity - \hat{E}_0)}\, .
\end{align} 
For $M$-photon lithography, $D \propto N^{-(M-1)}$. For $M>1$ this becomes a function that decreases polynomially with the number of pixels $N$.
 
In conclusion, we have presented a general method for establishing the quality of classical and quantum imaging procedures based on the Fisher information. The resulting Cram\'er-Rao bounds give better estimates of the resolution than the traditional Abbe and Rayleigh criteria, and can also take into account the imperfections and edge effects of the substrate. {Moreover, this technique can be used in imaging via higher-order photon correlations.} The Fisher information can be calculated using a {\sc povm} that models the substrate, or it can be inferred directly from the experimental data itself, circumventing the need for complex mathematical modelling. We found that a saturation limit of the substrate has relatively small effect on the imaging resolution. A utility function was defined that takes into account the resolution as well as the efficiency of creating the image. 

Part of this research was funded by the UK Engineering and Physical Sciences Research Council. The authors thank J. von Zanthier for valuable discussions on multi-photon imaging.

\end{document}